# The Area Signal-to-Noise Ratio: A Robust Alternative to Peak-Based SNR in Spectroscopic Analysis

Alex Yu[1,2], Huaqing Zhao[3], and Lin Z. Li[1]

[1]Britton Chance Laboratory of Redox Imaging, Department of Radiology, University of Pennsylvania Perelman School of Medicine, Philadelphia, PA
[2]Highland Park High School, Highland Park, NJ
[3]Center for Biostatistics and Epidemiology, Department of Biomedical Education and Data Science, Temple University School of Medicine, Philadelphia, PA

December 21, 2025

**Abstract**

In spectroscopic analysis, the peak-based signal-to-noise ratio (pSNR) is commonly used but suffers from limitations such as sensitivity to noise spikes and reduced effectiveness for broader peaks. We introduce the area-based signal-to-noise ratio (aSNR) as a robust alternative that integrates the signal over a defined region of interest, reducing noise variance and improving detection for various lineshapes. We used Monte Carlo simulations (n=2,000 trials per condition) to test aSNR on Gaussian, Lorentzian, and Voigt lineshapes. We found that aSNR requires significantly lower amplitudes than pSNR to achieve a 50% detection probability. Receiver operating characteristic (ROC) curves show that aSNR performs better than pSNR at low amplitudes. Our results show that aSNR works especially advantageously for broad peaks and could be extended to volume-based SNR for multidimensional spectra.

## 1 Introduction

The peak-based signal-to-noise ratio (pSNR) plays an important role in spectroscopic analysis, serving as a key metric for signal detection and quantification in fields like nuclear magnetic resonance (NMR), infrared (IR), and Raman spectroscopy [1-4]. However, pSNR has notable limitations, including high sensitivity to localized noise spikes and poor performance for distributed signals with broad lineshapes where the signal is distributed over multiple data points. pSNR measures the peak height by noise, i.e., standard deviation. This favors narrow spikes and is vulnerable to noise outliers. In contrast, spectroscopists routinely integrate peak area for quantification, but area is not used as the detection statistic itself in the field of spectroscopy.

We address this gap by formalizing an area-based SNR (aSNR) for signal detection by integrating the signal over a region of interest (ROI), thereby averaging noise across multiple points or bins and providing a more stable signal metric. Conceptually, aSNR also offers advantages in increased sensitivity for extended signals, making it better suited for real-world spectra with varying Lineshape widths. We found that aSNR achieves the same detection probability at substantially lower amplitudes than pSNR for broad peaks, while matching pSNR on narrow peaks where the number of bins is close to one. The aSNR metric significantly improves the signal detection and accuracy compared to pSNR using the Receiver

Operating Characteristic (ROC) analysis. We focus on 1D spectra with additive Gaussian noise and known linewidths to establish the advantage of aSNR. Section 2 defines the theory, Section 3 details the simulation protocol, Section 4 presents results, Section 5 discusses implications and limitations, and Section 6 is the conclusion.

## 2 Theory

At each point in the spectrum, the signal consists of the true signal plus random noise. We consider an additive noise function with an average of zero that is independent of other bins and evenly distributed with equal variance at each bin on a grid $\{x_i\}_{i=1}^{N}$ with spacing $\Delta x$. Mathematically, we represent this definition as

$$y_i = s_i + n_i, \qquad n_i \sim \mathcal{N}(0, \sigma^2) \tag{1}$$

where $y_i$ is the signal with noise, $s_i = s(x_i)$ is the true signal with a parametric lineshape centered at $x_0 = 0$ with peak height $h = s(0)$, $n_i$ is the noise with an average of zero and a standard deviation of $\sigma$. Unless otherwise noted, the peak location is assumed to be known for the purposes of defining the detection statistics.

The pSNR is defined as

$$\mathrm{pSNR} = \frac{|y(x_0)|}{\sigma} = \frac{|s(x_0) + n(x_0)|}{\sigma} \tag{2}$$

where $y(x_0)$ is the signal value at the peak position $x_0$, and the absolute value is utilized to handle inverted signal peaks. In search settings where the location is unknown and one scans all $N$ bins, the operational statistic is $\max_i |y_i|/\sigma$, but for the demonstrative purposes and proof of concept of this study, we (as will be shown in methods) will assume the peak center position is at $x_0=0$ without loss of generality. Common detection thresholds range from 3 to 5 for pSNR in the field of spectroscopy and imaging.

The aSNR is defined as

$$\mathrm{aSNR} = \frac{A_{\mathrm{ROI}}}{\sigma_{\mathrm{area}}} \tag{3}$$

where $\sigma_{\mathrm{area}}$ is the standard deviation of integrated noise and $A_{\mathrm{ROI}}$ is the integrated area over a predefined ROI. With the peak width $w$ based on a percentage threshold of height,

$$A_{\mathrm{ROI}} = \int_{-\frac{w}{2}}^{\frac{w}{2}} y(x)\, dx = \sum_{i \in \mathrm{ROI}}^{N_{\mathrm{ROI}}} y(x_i) \cdot \Delta x \tag{4}$$

where ROI is defined to be the set of bins with clean signal $s(x_i) \geq \eta \cdot s(0)$ ($\eta$ is a thresholding factor less than 1), $N_{\mathrm{ROI}}$ is the number of bins in the ROI, and $\Delta x$ or $dx$ is the bin width.

The denominator $\sigma_{\mathrm{area}}$ accounts for noise level when summing over $N_{\mathrm{ROI}}$ independent bins. The standard deviation of the sum scales as $\sqrt{N_{\mathrm{ROI}}}$, giving:

$$\sigma_{\text{area}} = \sigma \cdot \sqrt{N_{\text{ROI}}} \cdot \Delta x = \sigma \cdot \sqrt{\frac{w}{\Delta x}} \cdot \Delta x = \sigma \cdot \sqrt{w \cdot \Delta x} \tag{5}$$

The aSNR is now defined as

$$\text{aSNR} = \frac{\int_{-\frac{w}{2}}^{\frac{w}{2}} y(x)\, dx}{\sigma \cdot \sqrt{w \cdot \Delta x}} = \frac{\sum_{i \in \text{ROI}} y(x_i) \cdot \Delta x}{\sigma \cdot \sqrt{N_{\text{ROI}}} \cdot \Delta x} = \frac{\sum_{i \in \text{ROI}} y(x_i)}{\sigma \cdot \sqrt{N_{\text{ROI}}}} \tag{6}$$

Taking the ratio of aSNR over pSNR (equation 6 divided by equation 2), we obtain $\gamma$ as the improvement factor:

$$\gamma = \frac{\int_{-\frac{w}{2}}^{\frac{w}{2}} y(x)\, dx}{|y(0)|\sqrt{w \cdot \Delta x}} = \frac{\sum_{i \in \text{ROI}} y(x_i)}{|y(0)|\sqrt{N_{\text{ROI}}}} \tag{7}$$

where $w$ is the width, and $\Delta x$ is the discrete bin size.

We study Gaussian, Lorentzian, and Voigt line profiles, parameterized by peak height and width. The Gaussian lineshape is defined as follows:

$$s(x) = h \exp\left(-\frac{x^2}{2b_G^2}\right) \tag{8}$$

where $h$ is the peak height, $b_G$ is the width parameter, with area $\int s(x)\, dx = h\, b_G \sqrt{2\pi}$.

For a threshold level $\eta < 1$, the ROI width $w$ is determined by the condition $s(\pm w/2) = \eta \cdot h$, giving:

$$\exp\left(-\frac{w^2}{8b_G^2}\right) = \eta \quad \Rightarrow \quad w_G = 2b_G \sqrt{-2\ln(\eta)}$$

Substituting into the improvement factor (7) and evaluating the integral yields:

$$\gamma_G = \frac{\int_{-\frac{w}{2}}^{\frac{w}{2}} y_G(x)\, dx}{y(0) \cdot \sqrt{w \cdot \Delta x}} = \frac{\int_{-\frac{w_G}{2}}^{\frac{w_G}{2}} h \exp\left(-\frac{x^2}{2b_G^2}\right) dx}{h\sqrt{w \cdot \Delta x}} \tag{9}$$

This integral must be evaluated numerically for general $\eta$. For the commonly used half-maximum threshold ($\eta = 1/2$, FWHM), we obtain

$$\gamma_G \approx 1.24 \sqrt{\frac{b_G}{\Delta x}} \tag{10}$$

Using $w_G = 2b_G\sqrt{2\ln 2}$, we get that $\gamma_G \approx 0.81\sqrt{N_{\text{ROI}}}$.

The Lorentzian lineshape is defined as:

$$s(x) = \frac{h}{1 + (x/b_L)^2} \tag{11}$$

where $b_L$ is the width parameter and the area $\int s(x)\, dx = \pi h b_L$. The ROI width at threshold $\eta$ satisfies:

$$\frac{1}{1+(w/2b_L)^2} = \eta \quad \Rightarrow \quad w_L = 2b_L\sqrt{\frac{1-\eta}{\eta}} \tag{12}$$

The corresponding improvement factor is:

$$\gamma_L = \frac{\int_{-\frac{w}{2}}^{\frac{w}{2}} y_L(x)\,dx}{y(0) \cdot \sqrt{w \cdot \Delta x}} = \sqrt{2\frac{b_L}{\Delta x}} \frac{\arctan\left(\sqrt{\frac{1-\eta}{\eta}}\right)}{\sqrt[4]{\frac{1-\eta}{\eta}}} \tag{13}$$

At half-maximum ($\eta = 1/2$), this reduces to $w_L = 2b_L$, $\gamma_L = \frac{\pi}{4}\sqrt{2\frac{b_L}{\Delta x}} \approx 1.11\sqrt{\frac{b_L}{\Delta x}} = 0.78\sqrt{N_{\text{ROI}}}$. The Voigt profile is the convolution of Gaussian and Lorentzian components, normalized to peak height $h$. Its FWHM and area don't have simple closed-form solutions but can be computed numerically. Computational details will be provided in Methods.

All the above can be naturally extended to higher dimensions for multi-dimensional spectroscopy or image data. For example, for the detection of a spot on a 2D image, we could define a volume-based SNR (vSNR) by integrating the 2D signals over the spot area.

## 3 Methods

We systematically analyze the effectiveness of aSNR to pSNR in varying conditions of peaks and noises with theoretical modeling using MATLAB. We sample a 1D spectrum on a uniform grid $x \in [-20, 20]$ with a spacing of $\Delta x = 0.01$, yielding $N = 4001$ points. Observations follow Equation (1) with $\sigma = 1$ in all simulations.

### 3.1 Lineshapes

We generate peaks centered on $x_0 = 0$ using Gaussian (Equation *8*), Lorentzian (Equation *11*) and Voigt profiles. Lineshapes were matched by FWHM across all three types. For a target FWHM, we calculated: Gaussian parameter as $b_G = \text{FWHM}/(2\sqrt{2\ln(2)})$, Lorentzian parameter as $b_L = \text{FWHM}/2$, and Voigt parameters $b_{VG} = b_{VL}$ were determined numerically to match the target FWHM. The Voigt profile $s_V$ is numerically evaluated as the convolution of a unit peak Gaussian (width $b_G$) and a unit peak Lorentzian (half-width $b_L$). We discretize a symmetric span $[-S, S]$ with step $\delta$ (here $S = 20\max\{b_G, b_L\}$ and $\delta$ are typically $\Delta x/4$), form $v = \text{conv}(G, H)\,\delta$, normalize $v$ to the unit maximum, and then interpolate onto the simulation grid; the requested amplitude $a$ scales the result. The Voigt area is obtained by trapezoidal integration of the normalized kernel times $a$.

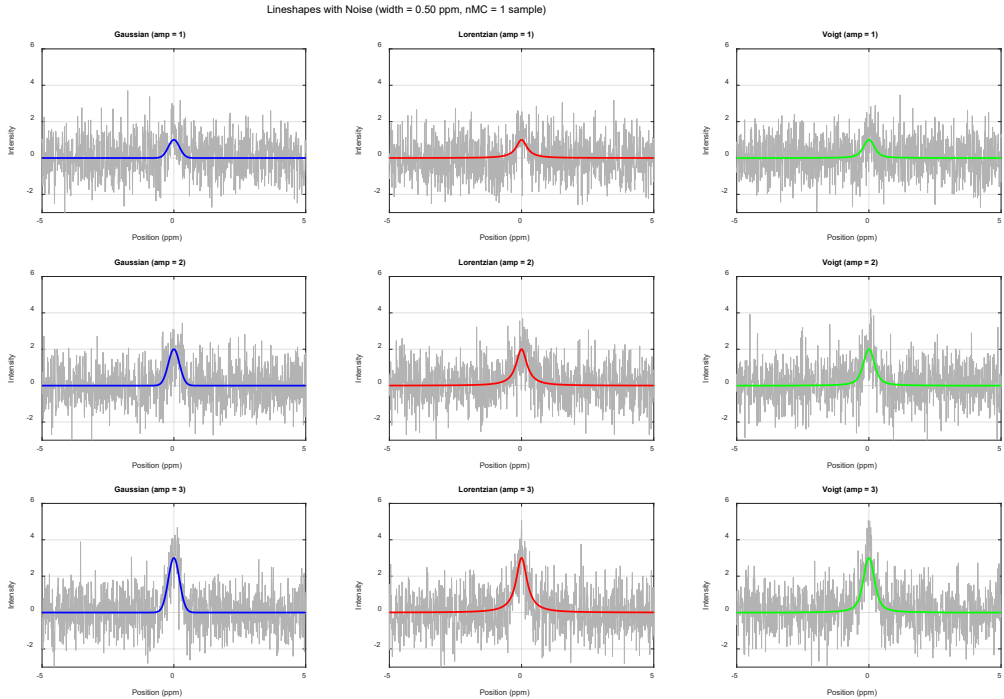

*Figure* **1:** *Representative examples of the three lineshapes used in simulations, embedded in Gaussian noise ($\sigma_{noise} = 1.0$) at amplitudes 1, 2, and 3. Gray traces show noisy backgrounds while colored lines (blue = Gaussian, red = Lorentzian, green = Voigt) show the underlying clean signal. Each row represents a different amplitude; columns represent different lineshape types. This figure illustrates the detection challenge at low amplitudes where signals are comparable to noise fluctuations.*

Figure *1* shows examples of each lineshape with noise at three different amplitudes. At amplitude 1, the peaks are hard to distinguish from background noise. At amplitude 2, the underlying signal shape becomes visible, but noise spikes are still similar in size to the signal. At amplitude 3, the clean lineshape is clearly visible above the noise. These examples illustrate the typical range of signal-to-noise conditions examined in our Monte Carlo simulations.

### 3.2  Region of Interest definition

The region of interest can be defined in different ways, such as the full-width at half-maximum (FWHM), full-width fifth maximum, or full-width tenth maximum, etc. Without losing generality, we defined the region of interest using the known full-width at half-maximum (FWHM) from the clean signal applied to the clean signal template. Using the clean signal for ROI thresholding excludes the effect of noise distortion on signal profiles. This simplification facilitates computer modeling without essentially affecting the comparison between pSNR and aSNR. For each lineshape, we identified all grid points where the signal exceeded half its maximum value:

$$\text{ROI} = \{i\, : \, s(x_i) \geq 1/2\, s_{\max}\}, \qquad N_{\text{ROI}} = |\text{ROI}| \tag{14}$$

We evaluated $s(x)$ on the simulation grid, identify all points where $s(x_i) \geq 0.5 s_{\max}$, and select the connected region containing the peak center $x_0$. The area was obtained by summing over these bins. The width of the bin $\Delta x$ cancels in normalization.

For 2D grids with coordinates $(x_i, y_j)$ and clean template $s(x, y)$, the ROI is the 2D half–max mask

$$\text{ROI}_{2D} = \{(i,j) : s(x_i, y_j) \geq 1/2\ s_{\max}\}, \qquad |\text{ROI}_{2D}|\ \text{pixels} \tag{15}$$

selecting only the region containing the peak center $(x_0, y_0)$. The 2D aSNR uses the standardized sum over this mask; the pixel area factor $(\Delta x\, \Delta y)$ cancels in the normalization. For each condition, we used the same region of interest (ROI) across all detection tests. We also tried other methods of defining the ROI (for example, using fixed window widths or amplitude thresholds on $y$) and found that the overall trends remained the same when the ROI widths were comparable.

### 3.3 Simulation design

We conducted three primary experiments to evaluate detection performance across different conditions. First, we varied the amplitude of the three lineshape profiles as we kept the width, noise, and bin count constant. Then, using representative amplitude values of 1, 2, and 3, keeping noise at 1, we vary the bin count. Last, we conduct receiver operating characteristic (ROC) analysis for representative amplitude-width pairs.

In the amplitude sweep, we varied peak height from 0 to 5 in steps of 0.5 while holding lineshape parameters constant. For each amplitude and lineshape combination, we computed pSNR and aSNR values across 2000 Monte Carlo trials (chosen to achieve stable mean and variance estimates with relative error ≤ 1%). We calculated mean and standard deviation for each statistic, then determined detection probabilities as the fraction of trials exceeding a threshold, e.g., 3 or 5. Linear interpolation between amplitude points yielded the critical amplitude at which each statistic achieved 50% detection probability.

In the width sweep, we varied the full-width at half-maximum from 1 to 50 bins (up to 100 bins in extended analyses) while holding amplitude fixed at representative values of 1, 2, and 3. We computed mean SNR values and detection probabilities across the same 2000 Monte Carlo trials by averaging the SNR values of each width/peak values.

### 3.4 Receiver Operating Characteristic Analysis

For the ROC analysis, we used binary hypothesis testing to compare detection performance at different signal levels. We generated $10^5$ Monte Carlo trial simulations under two conditions: the null hypothesis $H_0$, where spectra were noise-only trials with no signal present, and the alternative hypothesis $H_1$, for which true signal trials exist per trial. For the $H_0$ condition, spectra contained only Gaussian noise with $\sigma = 1$; for the $H_1$ condition, a signal Lineshape with specified amplitude was embedded in identical noise.

We evaluated three representative amplitudes (0.3, 0.4, and 0.5) that are below the critical amplitudes we found for pSNR in Section 3.3. For each amplitude, we tested both Gaussian Lineshapes ($b_G = 0.5$) and Lorentzian Lineshapes ($b_L = 0.5$), totaling eighteen experimental conditions. These amplitudes are all below pSNR's critical amplitude ( ~3.0 at threshold 3), where pSNR fails to detect signals, while aSNR maintains detection capability, spanning aSNR's critical amplitude of ~0.27-0.65 depending on threshold.

For each condition, we computed pSNR and aSNR values for all $10^5$ trials under both $H_0$ and $H_1$ hypotheses using the methodology described in Section 3.3. Detection threshold $\tau$ was swept from 0 to 10 in steps of 0.05, and for each threshold we calculated the true positive rate (TPR, fraction of $H_1$ trials exceeding $\tau$) and false positive rate (FPR, fraction of $H_0$ trials exceeding $\tau$). ROC curves were constructed by plotting TPR versus FPR across all thresholds. We computed the area under the curve (AUC) calculation via trapezoidal integration after sorting by FPR. For visualization of the underlying signal distributions, we generated a probability density distribution from amplitude 0.3 (Gaussian lineshape) under both hypotheses. This amplitude was selected since it demonstrates clear aSNR superiority (AUC>0.95) while showing pSNR's failure (AUC≈0.51), and falls between the critical amplitudes reported for aSNR (threshold 3) in Table *1*.

### 3.5 Two-Dimensional Extension

We extended the analysis to two-dimensional spectra using a coarser grid ($\Delta x = 0.1$) spanning $x, y \in [-5,5]$ to manage memory requirements. This produced 101 × 101 spatial grids with corresponding three-dimensional noise arrays of dimensions N × N × nMC. We implemented 2D Gaussian, Lorentzian, and Voigt profiles as separable products of their 1D versions and computed volume-based SNR (vSNR) using the same ROI approach extended to two dimensions.

## 4 Results

We compared area-based SNR (aSNR) to peak-based SNR (pSNR) for one-dimensional and two-dimensional spectroscopic lineshapes under controlled noise conditions. Monte Carlo simulations revealed that aSNR detects signals much more effectively than pSNR, especially for broad peaks at high detection thresholds.

## 4.1 Detection Probability Curves

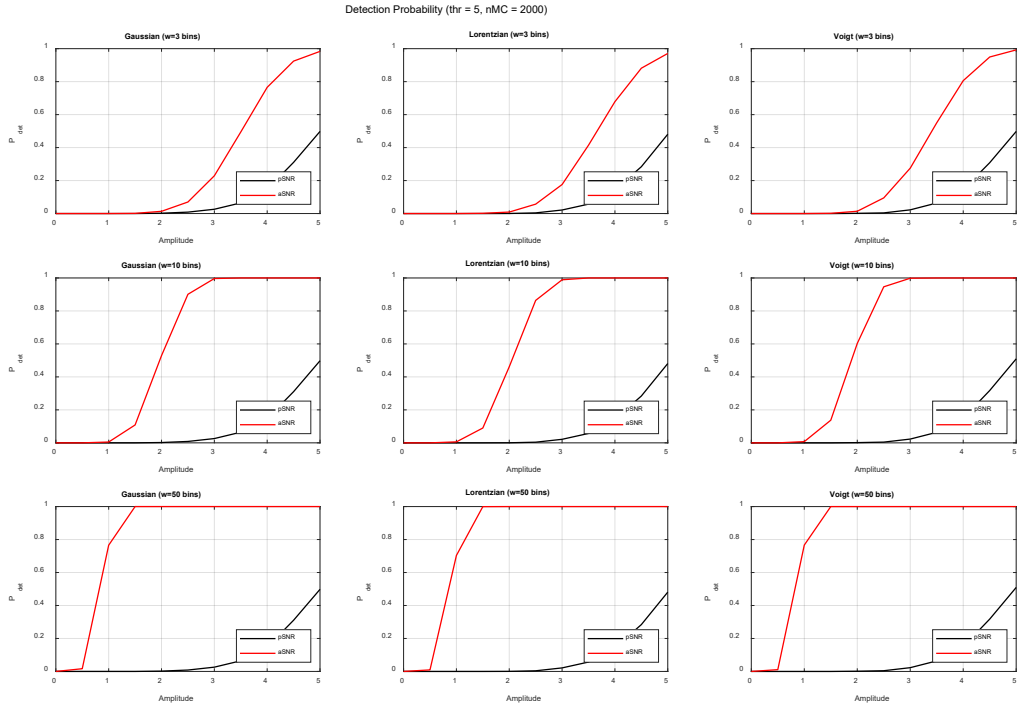

**Figure 2:** *Detection probability as a function of amplitude for Gaussian, Lorentzian, and Voigt Lineshapes at threshold $\tau = 5$ with $n_{MC} = 2000$ Monte Carlo trials. aSNR (red) achieves 50% detection at significantly lower amplitudes than pSNR (black) for all lineshapes.*

Figure 2 presents detection probability as a function of amplitude for threshold 5 in a 3×3 grid, with rows representing three different peak FWHM (3, 10, and 50 bins) and columns showing the three lineshape types (Gaussian, Lorentzian, Voigt). This layout makes it easy to see how aSNR's advantage changes with different lineshape types and peak widths. As expected, all curves show detection probability rising from near zero to high values as amplitude increases.

The aSNR curves (red) rise much more steeply than pSNR curves (black) in all cases, reaching over 95% detection at amplitudes where pSNR stays below 50 %. However, aSNR's advantage strongly depends on peak width. At narrow widths (eg. 3 bins), both methods render similar performances. As width increases to 10 and then 50 bins, the aSNR curves get steeper and shift to lower critical amplitudes, while pSNR curves stay about the same. This demonstrates that aSNR's advantage scales directly with the number of bins available for spatial integration. The $\sqrt{N_{\text{ROI}}}$ noise reduction becomes more powerful as the ROI encompasses more independent spectral points.

Comparing across lineshapes (columns), Gaussian profiles show the most gradual aSNR transitions because their signal concentrates near the peak center. Lorentzian and Voigt profiles have broader wings, so they show even steeper aSNR transitions at matched widths since their extended tails add more signal to the integrated ROI. This lineshape dependence becomes more obvious at wider peak widths (bottom row), where Voigt profiles achieve near-perfect detection at the lowest amplitudes.

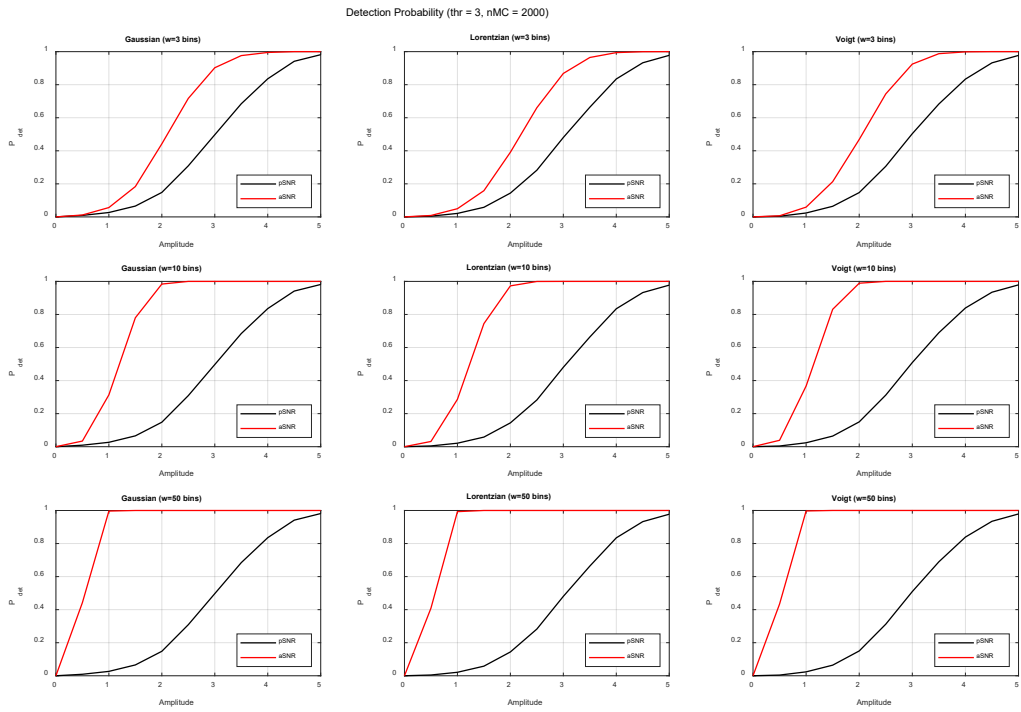

*Figure* **3:** *Detection probability as a function of amplitude at threshold* $\tau = 3$. *Both metrics perform better at this lower threshold, but aSNR maintains substantial advantages with improvement factors ranging from 1.9-13×.*

At the lower detection threshold of 3 (Figure *3*), both metrics performed better overall, but aSNR maintained a substantial advantage over pSNR.

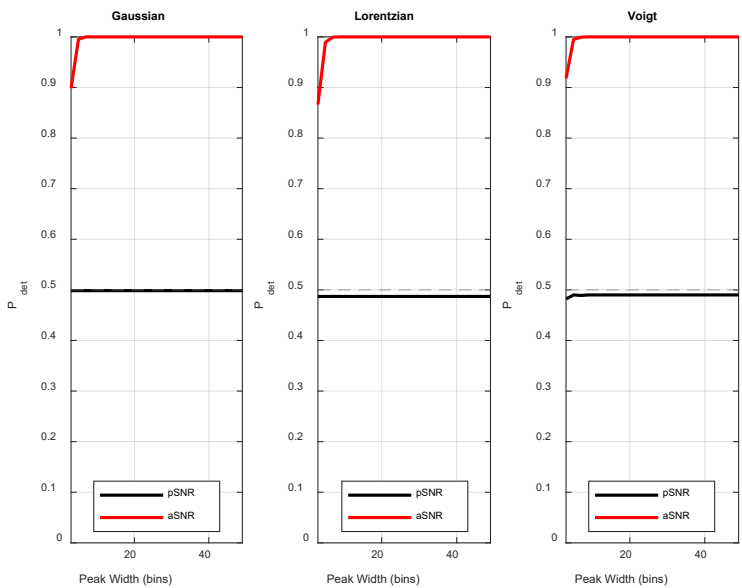

*Figure* **4:** *Detection probability as a function of peak width at fixed amplitude and threshold $\tau = 5$. For narrow peaks (width < 5 bins), both metrics converge, while aSNR demonstrates superior performance for broader features.*

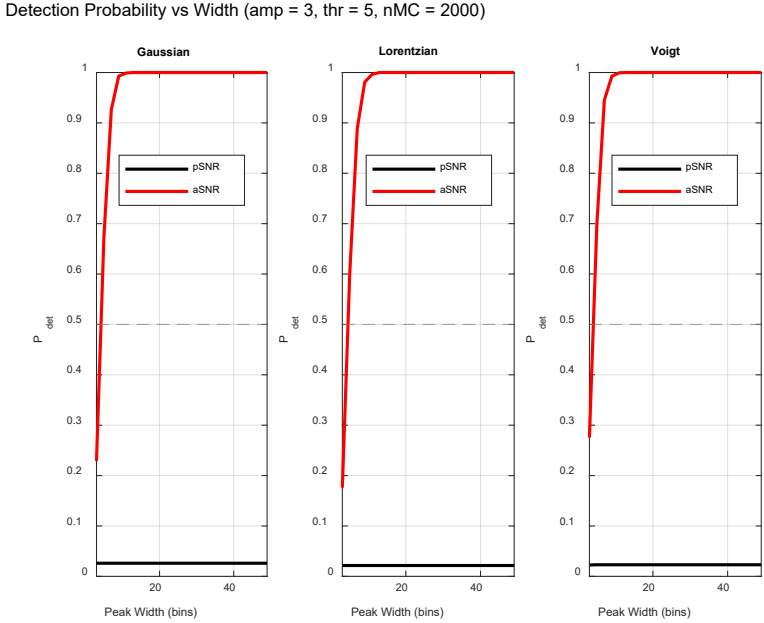

*Figure* **5:** *Detection probability versus peak width at threshold $\tau = 3$, showing similar convergence behavior at narrow widths with maintained aSNR advantage for broader peaks.*

Figures *4* and *5* show the detection probability as a function of the FWHM when the amplitude is set at 3. Figure *4* shows that, for the detection threshold of 5, aSNR detection probability increases rapidly with the peak width, quickly reaching 1 before a FWHM of 10 bins, while the peak is hardly detected using pSNR, with a detection probability less than 0.05. This is consistent with majority peak heights should be less than the threshold of 5 when the amplitude is 3. *5* compares pSNR and aSNR at a lower commonly used threshold of 3, where our pSNR simulations accordingly achieve a detection rate of ∼ 0.5. However, at the SNR threshold of 3, aSNR reaches a detection probability of 1 at a FWHM of less than ∼ 5 bins, which is lower in bin count than the SNR threshold of 5, as expected.

### 4.2    Critical Detection Amplitudes

To quantify detection performance, we extracted the amplitude at which each metric achieved 50% detection probability. With Gaussian noise, this 50% point indicates that the mean signal SNR value equals the detection threshold, where half of the trials exceed the threshold and half fall below. Each amplitude sweep was repeated ten times with independent noise, and the mean critical amplitudes with standard deviations were calculated across these repetitions. Table 1 summarizes these critical amplitudes at thresholds of 3 and 5, showing critical amplitudes for 50% detection probability across line widths. The improvement factor is the ratio of respective critical amplitudes for pSNR to aSNR.

**Table 1** *Sampled critical amplitudes for 50% detection probability across bin widths (n=10 trials)*

| Lineshape | Width (Bins) | Threshold | pSNR Crit. Amp | aSNR Crit. Amp | Improvement Factor |
|---|---|---|---|---|---|
| Gaussian | 3 | 3 | 3.00245 ± 0.03165 | 2.10775 ± 0.01453 | 1.42× |
| Gaussian | 10 | 3 | 3.00245 ± 0.03165 | 1.19174 ± 0.00594 | 2.52× |
| Gaussian | 50 | 3 | 3.00245 ± 0.03165 | 0.54450 ± 0.00771 | 5.51× |
| Lorentzian | 3 | 3 | 2.99942 ± 0.03084 | 2.18799 ± 0.01383 | 1.37× |
| Lorentzian | 10 | 3 | 2.99942 ± 0.03084 | 1.22905 ± 0.00732 | 2.44× |
| Lorentzian | 50 | 3 | 2.99942 ± 0.03084 | 0.56955 ± 0.00754 | 5.27× |
| Voigt | 3 | 3 | 3.03123 ± 0.02673 | 2.05431 ± 0.02814 | 1.48× |
| Voigt | 10 | 3 | 3.01778 ± 0.02598 | 1.15713 ± 0.00466 | 2.61× |
| Voigt | 50 | 3 | 3.01752 ± 0.02612 | 0.55448 ± 0.00802 | 5.44× |
| Gaussian | 3 | 5 | 5.00251 ± 0.03012 | 3.51192 ± 0.01887 | 1.42× |
| Gaussian | 10 | 5 | 5.00251 ± 0.03012 | 1.97186 ± 0.01223 | 2.54× |
| Gaussian | 50 | 5 | 5.00251 ± 0.03012 | 0.81850 ± 0.00394 | 6.11× |
| Lorentzian | 3 | 5 | 4.99750 ± 0.03170 | 3.63929 ± 0.01370 | 1.37× |
| Lorentzian | 10 | 5 | 4.99750 ± 0.03170 | 2.05396 ± 0.01225 | 2.43× |
| Lorentzian | 50 | 5 | 4.99750 ± 0.03170 | 0.84461 ± 0.00495 | 5.92× |
| Voigt | 3 | 5 | 5.04313 ± 0.02504 | 3.41168 ± 0.01969 | 1.48× |
| Voigt | 10 | 5 | 5.01840 ± 0.02466 | 1.89390 ± 0.00530 | 2.65× |
| Voigt | 50 | 5 | 5.01800 ± 0.02502 | 0.82554 ± 0.00387 | 6.08× |

Across all conditions, aSNR consistently reaches 50% detection probability at much lower amplitudes than pSNR, and the improvement gets more apparent as peaks get wider. The advantage is largest for broad peaks (50 bins), with all three lineshapes performing similarly when FWHM is matched (improvement factors ranging from 5.3× to 5.5× at threshold 3, and 5.9× to 6.1× at threshold 5). This scaling shows that aSNR's advantage comes from integrating over multiple independent noise samples. At the SNR threshold of 5, the improvement factor approximately agrees with the theoretical expectation (Eqs 10 and 13).

## 4.3 Width-Dependent Performance

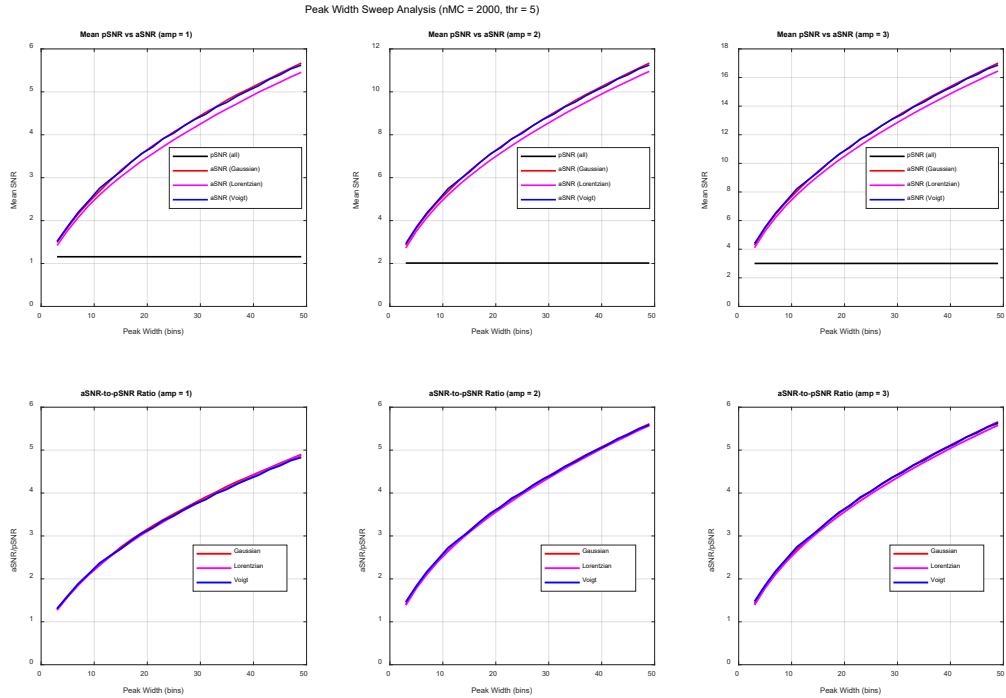

Figure **6:** *Peak width sweep analysis at fixed amplitudes (1, 2, and 3). Top row: Mean SNR versus peak width (1-50 bins). pSNR remains constant (horizontal lines) while aSNR increases monotonically with width. Bottom row: aSNR/pSNR ratio, reaching 5-6× at 50 bins depending on lineshape.*

The width sweep analysis (Figure 6) further reveals how the aSNR advantage scales with peak width. The top panels show mean SNR versus peak width (1 to 50 bins) at three fixed amplitudes (1, 2, and 3). Peak-based SNR remains essentially constant across all widths, as expected–peak height is independent of width for fixed amplitude in these parameterizations. In contrast, area-based SNR increases monotonically with width for all lineshapes, with distinct curves for Gaussian (red), Lorentzian (magenta), and Voigt (blue).

The aSNR/pSNR ratio (bottom panels) quantifies this advantage. Starting near 1 for narrow peaks (1-2 bins, where the ROI contains effectively a single point), the ratio increases approximately with a square root relationship with width. At 50 bins, the Gaussian lineshape (represented in red), the Ratio reaches ∼ 4.82× at amp=1, ∼5.58× for amp=2, and ∼ 5.64 × for amp=3. For the Lorentzian lineshape (represented in magenta), the ratio reaches ∼4.73× at amp=1, ∼5.45× at amp=2, and ∼5.49× at amp=3. The Voigt lineshape (in blue) reaches ∼ 4.87 × at amp=1,∼ 5.60 × at amplitude 2, and ∼5.63× at amp=3. All three lineshapes perform similarly when FWHM is matched, with Gaussian and Voigt showing very similar performance, and Lorentzian performing within 3% of those two.

Notably, at amplitude 3 (right column), pSNR exceeds the threshold of 5 (indicated by horizontal reference line in top panels) for all widths, leading to high detection probability regardless of metric.

However, at lower amplitudes (1 and 2), only aSNR maintains sufficient sensitivity as width increases, explaining the dramatic detection improvements observed in Section 4.1.

## 4.4   Receiver Operating Characteristic Analysis

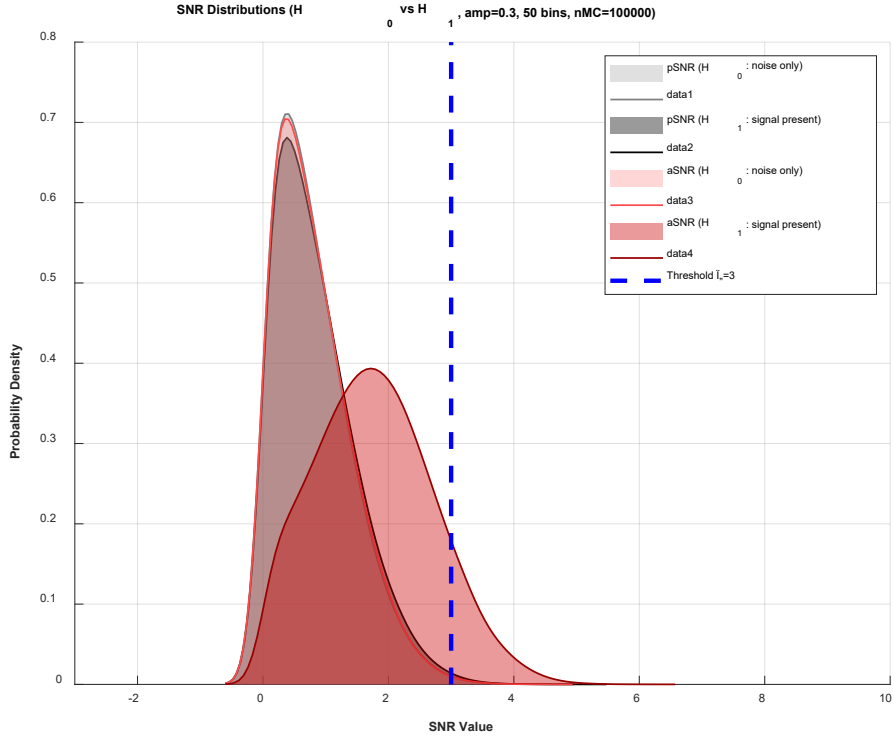

Figure **7:** *Probability density distributions of pSNR and aSNR under null hypothesis $H_0$ (noise-only, light shading) and alternative hypothesis $H_1$ (signal-present, dark shading) for a Gaussian lineshape at amplitude 0.3 with $n_{MC} = 10^5$ Monte Carlo trials.*

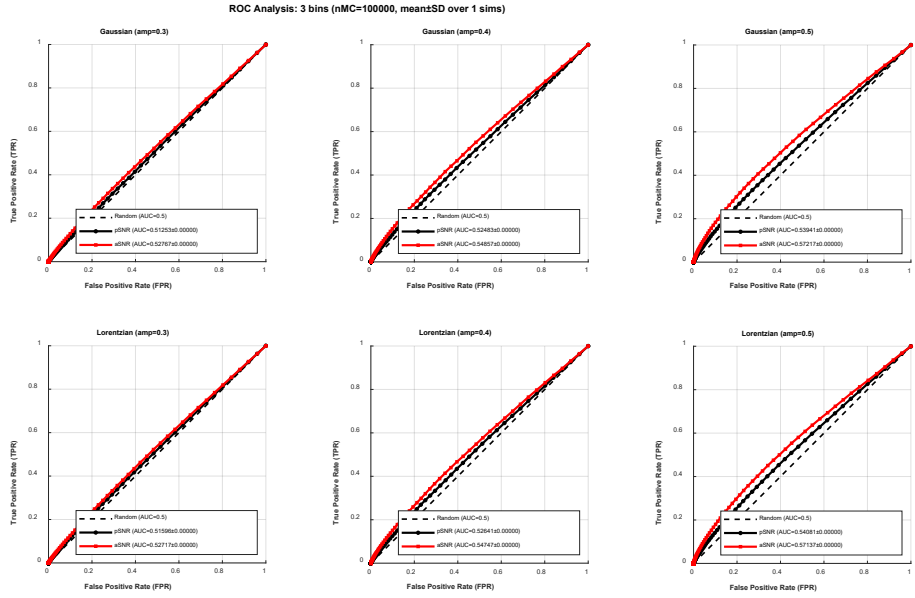

Figure **8:** *ROC curves comparing aSNR (red) and pSNR (black) detection performance for Gaussian (top row) and Lorentzian (bottom row) lineshapes at 3 bins across three amplitudes (0.3, 0.4, 0.5) with $n_{MC} = 10^5$ trials. Area under the curve (AUC) values are shown in legends. At narrow widths, both metrics show limited discrimination, with aSNR achieving modest improvements (Gaussian: AUC=0.56–0.66; Lorentzian: AUC=0.55–0.63) over pSNR (AUC=0.51–0.54).*

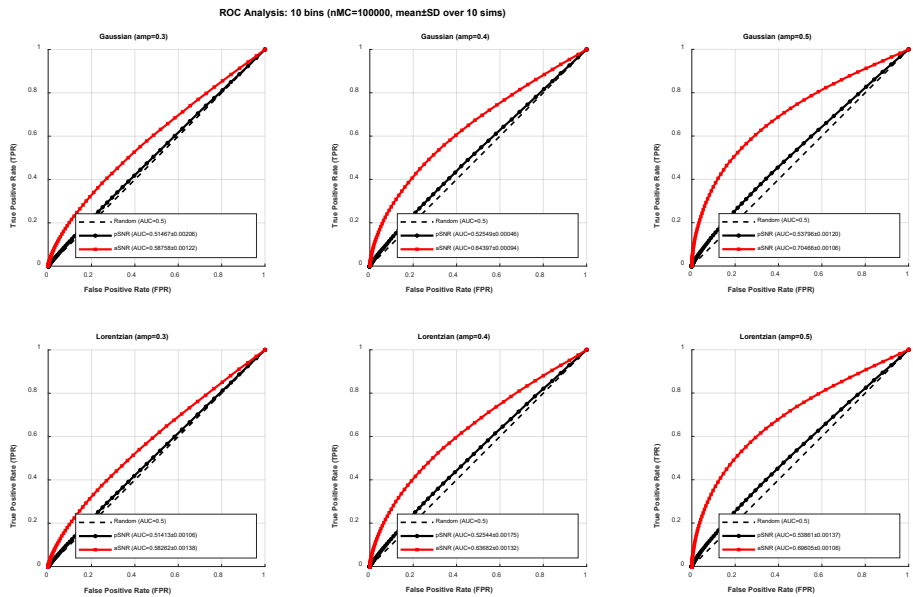

Figure **9:** *ROC curves at 10 bins with $n_{MC} = 10^5$ trials. At this intermediate width, aSNR demonstrates clear advantages (Gaussian: AUC=0.68–0.85; Lorentzian: AUC=0.65–0.81) while pSNR remains near random classification (AUC=0.51–0.54).*

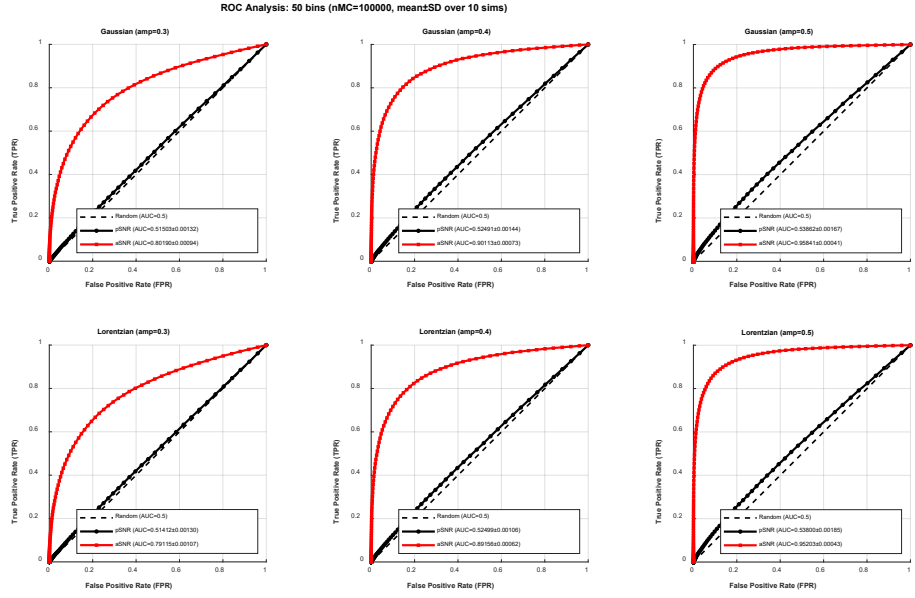

Figure **10:** *ROC curves at 50 bins with $n_{MC} = 10^5$ trials. At this broad width, aSNR achieves excellent-to-perfect discrimination (Gaussian: AUC=0.94–1.00; Lorentzian: AUC=0.91–1.00) while pSNR remains ineffective (AUC=0.51–0.54).*

**Table 2** *Summary of ROC analysis results: AUC values and improvement factors across bin widths (n=10 trials)*

| Lineshape | Bin Width | Amplitude | pSNR AUC | aSNR AUC | Improvement |
|---|---|---|---|---|---|
| Gaussian | 3 | 0.3 | 0.51372 ± 0.00159 | 0.52917 ± 0.00174 | 2.99% |
| Gaussian | 3 | 0.4 | 0.52469 ± 0.00080 | 0.54965 ± 0.00103 | 4.76% |
| Gaussian | 3 | 0.5 | 0.53817 ± 0.00069 | 0.57507 ± 0.00117 | 6.86% |
| Gaussian | 10 | 0.3 | 0.51413 ± 0.00201 | 0.58709 ± 0.00139 | 14.19% |
| Gaussian | 10 | 0.4 | 0.52480 ± 0.00129 | 0.64365 ± 0.00112 | 22.64% |
| Gaussian | 10 | 0.5 | 0.53933 ± 0.00147 | 0.70500 ± 0.00131 | 30.72% |
| Gaussian | 50 | 0.3 | 0.51466 ± 0.00089 | 0.80190 ± 0.00078 | 55.79% |
| Gaussian | 50 | 0.4 | 0.52451 ± 0.00160 | 0.90070 ± 0.00046 | 71.75% |
| Gaussian | 50 | 0.5 | 0.53865 ± 0.00190 | 0.95830 ± 0.00037 | 77.92% |
| Lorentzian | 3 | 0.3 | 0.51494 ± 0.00135 | 0.52666 ± 0.00112 | 2.28% |
| Lorentzian | 3 | 0.4 | 0.52459 ± 0.00117 | 0.54571 ± 0.00113 | 4.03% |
| Lorentzian | 3 | 0.5 | 0.53842 ± 0.00140 | 0.57055 ± 0.00069 | 5.97% |
| Lorentzian | 10 | 0.3 | 0.51455 ± 0.00118 | 0.58293 ± 0.00126 | 13.29% |
| Lorentzian | 10 | 0.4 | 0.52522 ± 0.00170 | 0.63710 ± 0.00118 | 21.30% |
| Lorentzian | 10 | 0.5 | 0.53853 ± 0.00143 | 0.69574 ± 0.00111 | 29.20% |
| Lorentzian | 50 | 0.3 | 0.51454 ± 0.00084 | 0.79147 ± 0.00104 | 53.80% |
| Lorentzian | 50 | 0.4 | 0.52558 ± 0.00114 | 0.89153 ± 0.00083 | 69.61% |
| Lorentzian | 50 | 0.5 | 0.53765 ± 0.00103 | 0.95223 ± 0.00040 | 77.12% |

ROC analyses were performed to compare aSNR to pSNR by computing the TPR and FPR under both H0 and H1 conditions. Before plotting the ROC curves, we show Figure *7* to illustrate how aSNR's detection capability scales systematically with peak width through the underlying probability distributions. At amplitude 0.3, these three cases span the range from narrow peaks where spatial integration provides modest benefit to broad peaks where aSNR achieves dramatic superiority over pSNR. Specifically, the pSNR distributions (gray) exhibit nearly complete overlap between $H_0$ (mean=0.798) and $H_1$ (mean=0.835), with a mean separation of only 0.037—less than 5% of the noise standard deviation. In contrast, aSNR distributions (red) show clear separation between $H_0$ (mean=−0.001) and $H_1$ (mean=2.638), with a mean separation of 2.64 standard deviations—a 70-fold improvement in separation. The dashed blue line indicates the standard detection threshold $\tau = 3$. This distribution separation explains aSNR's superior detection capability through spatial integration, which fundamentally transforms signal extraction by averaging noise across the FWHM region.

Figures *8*, *9*, *10* presents representative ROC curves with widths of 3, 10, 50 bins, respectively, demonstrating aSNR's detection advantage across three different amplitudes by varying signal detection thresholds through Monte Carlo simulations. Each ROC simulation was repeated ten times with independent noise, and the mean AUC with standard deviations were calculated across these repetitions as shown in Table 2. Examining performance across amplitudes (left to right in each row) reveals systematic improvement for both metrics as signal strength increases. The aSNR advantage is most exemplified at 50 bins. For example, for the Gaussian lineshape at FWHM=50 bins, aSNR achieves excellent discrimination across all amplitudes shown, with AUC values increasing from 0.802 (amp=0.3), 0.901 (amp=0.4), and 0.958 (amp=0.5), demonstrating progression from good to excellent to near-perfect detection from 0.3 to 0.5 peak height. In contrast, pSNR shows minimal improvement, remaining near random classification with AUC values of 0.515, 0.525, and 0.539, respectively, at 50 bins. This represents a 56 - 78% improvement in AUC across all amplitudes tested, calculated from taking the ratio of the AUC values.

Table 2 shows 18 experimental conditions and how aSNR's peak detection advantage scales with both amplitude and peak width across all of them. For each lineshape-width-amplitude combination, we report AUC values for both metrics and calculate the relative improvement as $(AUC_{aSNR} - AUC_{pSNR})/(AUC_{pSNR})$, representing the percentage increase in the capability of distinguishing signal from noise.

This (Table 2) shows several important trends. First, aSNR's advantage increases significantly with peak width when amplitude and lineshape are held constant. For Gaussian peaks at amplitude 0.3, the improvement factor goes from 2.99% (3 bins) to 14.19% (10 bins) to 55.79% (50 bins). This follows the $\sqrt{N_{ROI}}$ noise reduction scaling—wider peaks give more data points to integrate over, making aSNR's statistical power stronger. At 50 bins, aSNR reaches excellent discrimination (AUC >0.9) at all amplitudes for both Gaussian and Lorentzian lineshapes, while pSNR stays near random performance no matter the width.

Second, amplitude scaling shows that higher signal strengths improve both metrics but disproportionately benefit aSNR. Within each width category, AUC values increase monotonically with amplitude for aSNR (e.g., Gaussian 50 bins: 0.802, 0.901, 0.958 for amplitudes 0.3, 0.4, 0.5), approaching perfect discrimination at the highest amplitudes. However, the relative improvement factor can decrease at very high amplitudes (e.g., Gaussian 50 bins: 55.79%, 71.75%, 77.92%), possibly because aSNR approaches the ceiling of AUC=1.0, whereas pSNR has more room for improvement with increasing amplitude.

Third, **lineshape comparison** reveals consistent patterns. Gaussian profiles generally achieve slightly higher AUC values than Lorentzian at matched FWHM (e.g., at 10 bins, amp=0.3: Gaussian 0.587 vs Lorentzian 0.583), consistent with Gaussian's more compact signal distribution concentrating more energy within the FWHM ROI. However, both lineshapes demonstrate substantial aSNR advantages, with the improvement factors ranging from 2-78% across different conditions.

Finally, at narrow widths, both methods converge, confirming our theoretical predictions. At 3 bins, improvement factors are modest (roughly 2-7%) because the ROI contains only a few data points, so spatial integration doesn't help as much. Both metrics perform similarly in this case, as shown in Figure 8, confirming that aSNR's advantage comes specifically from integrating over extended spatial regions.

## 4.5 Two-Dimensional Extension

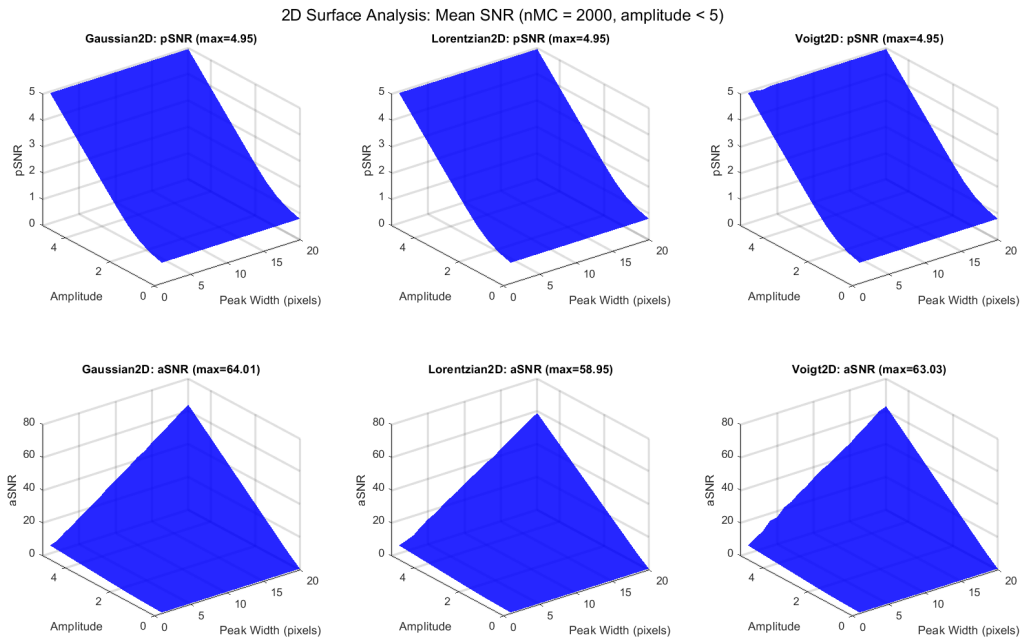

Figure **11:** *Two-dimensional surface plots showing mean pSNR (top row) and mean aSNR (bottom row) as functions of peak width (1-20 pixels) and amplitude (0-5). The blue surfaces represent mean values. aSNR surfaces form pyramids reaching 60-65, representing a 12-13× enhancement over pSNR.*

Extending the analysis to 2D lineshapes produced even more dramatic results. Figure *11* displays surface plots of mean pSNR (top row) and mean aSNR (bottom row) as functions of both peak width (1-20 pixels) and amplitude (0-5).

The pSNR surfaces (top row) are relatively flat planes, with values remaining below 10 across the entire parameter space. In stark contrast, the aSNR surfaces (bottom row) form pyramids reaching approximately 64 at the highest amplitudes and widths. For the Gaussian lineshape, the max pSNR value was 4.94, compared to the max aSNR value of 63.94. This represents a 12.95× enhancement. For the Lorentzian lineshape, the maximum pSNR value was 4.94, whereas the maximum aSNR value was 58.87, resulting in an 11.92× rate of increase. For the Voigt lineshape, the maximum pSNR value was 4.94, the maximum aSNR value was 62.98, totalling to a 12.75× enhancement over pSNR.

The pyramidal geometry of aSNR surfaces reflects the quadratic growth in ROI size with increasing 2D width (area scales as width$^2$), compounding the $\sqrt{N_{\text{ROI}}}$ noise reduction benefit. All three lineshape families (Gaussian2D, Lorentzian2D, Voigt2D) exhibit similar surface morphology, with all three lineshapes achieving similar peak aSNR values when FWHM is matched (within 8% of each other: Gaussian 12.95×, Lorentzian 11.92×, Voigt 12.75×).

# 5 Discussion

## 5.1 Advantage of aSNR over pSNR

The pSNR metric for signal detection is biased toward sharp features, overestimating noise spikes and underestimating broad signals. In this study, we formalize an area-based detection statistic (aSNR) that sums the signal over a region of interest and normalizes by the standard deviation of the sum of the noise in the region. We compared aSNR and pSNR using Monte Carlo simulations with different amplitudes and widths for Gaussian, Lorentzian, and Voigt lineshapes. aSNR consistently outperformed pSNR in all conditions we tested. The advantage increases with peak width and dimensionality, but becomes similar to pSNR for narrow peaks where area integration does not help as much. At the same detection threshold, aSNR has higher detection probability (Figures 2 and 3) and lower critical detection amplitude (Table 1). In other words, for broad peaks, aSNR reaches a given detection probability at much lower amplitudes than pSNR. For narrow peaks with only a few bins, the two methods perform similarly (Figures 4-6 and 11).

The ROC analysis confirms that aSNR detects signals better than pSNR across the full range of conditions relevant to spectroscopic applications. The ROC curves (Figures *8*, *9*, *10*) show that spatial integration creates better separation between noise and signal as $N_{ROI}$ increases, matching our theoretical $\sqrt{N_{ROI}}$ predictions. Table 2 quantifies how this advantage scales systematically with both peak width and signal amplitude, offering clear guidance on when aSNR provides significant benefits: broad peaks (10 bins or more) at low-to-moderate signal strengths where pSNR fails but aSNR achieves excellent discrimination. The SNR density distribution plot (Figure *7*) reveals the mechanism, i.e., spatial integration creates clear separation between noise and signal distributions. Together, these results confirm that aSNR transforms challenging detection regimes where pSNR fails into scenarios with excellent discrimination capability, offering spectroscopists a practical path to detecting signals at substantially lower amplitudes than previously achievable.

## 5.2 Threshold Dependence

Comparing results at signal detection thresholds 3 and 5 reveals that aSNR's relative advantage, i.e., the improvement factor varies with detection criteria at relatively larger widths (50 bins) (Table 1). While the Gaussian and Lorentzian improvement factors decrease slightly, Voigt improvement factor increases. This threshold dependence likely reflects the interplay between peak shape and the statistical power of area integration. More investigation is needed to understand this variation.

### 5.3 Lineshape Dependence

The improvement factor $\gamma$ (Equation 7) measures how efficiently each lineshape converts peak height to integrated signal within the FWHM region. At the half-maximum threshold, Lorentzian profiles exhibit $\gamma_L \approx 0.78\sqrt{N_{\text{ROI}}}$ compared to Gaussian's $\gamma_G \approx 0.81\sqrt{N_{\text{ROI}}}$, with Voigt as intermediate. As predicted by theory, Gaussian achieves a slightly higher improvement factor (0.81 vs. 0.78, 3.8% difference), which corresponds to simulations, reflecting Gaussian's more concentrated signal distribution within the FWHM region (76% compared to 50% for the Lorentzian lineshape). The theoretical predictions (Equations 9 and 13) are useful estimates, but the observed simulation ratios are somewhat higher, likely due to discrete sampling effects. The ROC analysis (Table 2) shows similar patterns, with Gaussian achieving slightly higher AUC values than Lorentzian at matched FWHM (e.g., 0.802 vs 0.791 at 50 bins, amp=0.3) reflecting Gaussian's higher signal density within the compact FWHM region.

When lineshapes are compared at matched FWHM, all three profiles show nearly identical performance. At 50 bins with threshold 5, improvement factors are: Gaussian 6.11×, Lorentzian 5.92×, and Voigt 6.08×—all within 3% of each other. This validates our theoretical prediction that γ coefficients are similar (γ_G ≈ 0.81, γ_L ≈ 0.78, γ_V ≈ 0.80). The small differences reflect the balance between peak concentration (favoring Gaussian) and extended tails (favoring Lorentzian), with Voigt exhibiting intermediate characteristics.

### 5.4 Limitations and Future Directions

Several statistical limitations warrant consideration. First, our Monte Carlo approach assumes independence between trials, which holds for simulated data but may require verification in experimental settings with correlated noise. Third, the ROC analysis assumes that the underlying distributions are reasonably approximated by simulation model; violations of these distributional assumptions could affect optimal threshold selection. In this study, we assume an ROI that is clearly defined by the known lineshape. The aSNR metric may conflate sharp spikes with flat curves if ROI is poorly defined. In the future, we may test aSNR using adaptive ROI selection when the lineshape is not a priority. We may also extend the use of aSNR to real spectroscopic datasets.

## 6 Conclusion

ASNR offers a robust alternative to pSNR that improves detection in spectroscopic analysis, especially for broad signals. Its integration-based approach has both conceptual and practical benefits that support its use in spectroscopic signal analysis.

## 7 Acknowledgements

We acknowledge funding from the National Institutes of Health (NIH) grant R01-CA277037.

# 8 References

[1] G. McGibney, M.R. Smith, An unbiased signal-to-noise ratio measure for magnetic resonance images, Medical Physics 20(4) (1993) 1077-1078.
[2] R.G. Brereton, Chemometrics and Statistics | Signal Processing, in: P. Worsfold, C. Poole, A. Townshend, M. Miró (Eds.), Encyclopedia of Analytical Science (Third Edition), Academic Press, Oxford, 2019, pp. 510-516.
[3] V. Saptari, Signal-to-Noise Ratio, in: Fourier-Transform Spectroscopy Instrumentation Engineering, SPIE2003, pp. 11-17.
[4] R.L. McCreery, Signal-to-Noise in Raman Spectroscopy, in: J.D. Winefordner, R.L. McCreery (Eds.), Raman Spectroscopy for Chemical Analysis2000, pp. 49-71.